\documentstyle[12pt]{article}

\def\be{\begin{equation}}
\def\ee{\end{equation}}
\def\bea{\begin{eqnarray}}
\def\eea{\end{eqnarray}}

\begin{document}

\pagestyle{plain} 

\def\e{{\rm e}}
\def\cs{\frac{1}{(2\pi\alpha')^2}}
\def\CV{{\cal{V}}}
\def\haf{{\frac{1}{2}}}
\def\tr{{\rm Tr}}

\baselineskip .55 cm

\begin{center}
{\Large {\bf Do Quarks Obey D-Brane Dynamics?}}

\vspace{.4cm}

Amir H. Fatollahi

{\it Institute for Studies in Theoretical Physics and Mathematics (IPM),
P.O.Box:19395-5531, Tehran, IRAN.}\\ 
\vskip .2 in 
{\sl fath@theory.ipm.ac.ir}

\vskip .2 in 
\begin{abstract}
The potential between two D0-branes at rest is calculated to be a linear. 
Also the potential between two fast decaying
D0-branes is found in agreement with phenomenology.
\end{abstract}
\end{center}

\vskip .05 in 

\section{Introduction}
In the last few years our understanding about string theory is
changed dramatically; a stream which is called "second string
revolution" \cite{vafa}. The scope of this stream is presentation of
a unified string theory as a fundamental theory of the known interactions.
One of the most applicable tools in the above program are D$p$-branes \cite
{Po1,Po2}. It was conjectured, and
is confirmed by various tests, that these objects can be considered as a
perturbative representation of nonperturbative(BPS) charged solutions of the
low energy of superstring theories.

On the other hand, the idea of string 
theoretic description of gauge theories is an old idea 
\cite{Po3} \cite{Polya}. Despite of the years that passed on this idea, it
is also activating different research in theoretical physics 
\cite{Polyakov} \cite{Pes}. 

It has been known for long time that hadron-hadron scattering processes
have two different behaviors depending on the amount of the momentum 
transfers \cite{Close}. At the large momentum transfers interactions appear
as interactions between the hadron constituents, partons or quarks, 
and some qualitative similarities to electron-hadron scattering emerge.
At high energies and small momentum transfers Regge trajectories are 
exchanged, the same which was the first motivation for the string picture 
of strong interactions 
\footnote{Regge behavior recently is used
for fitting the experimental data with a considerable success \cite{DL}}.
Besides the good fitting between Regge 
trajectories and the mass of strong bound states has remained 
unexplained yet.

Deducing the above partially different observations 
from a unified picture is attractive
and it is tempting to search for the application of the 
recent string theoretic progresses in this area.

In this way one finds D$p$-branes good tools. As the first step
we try to extract some known results from the dynamics of 
D$p$-branes. It is found that the
potential between static D0-branes is a linear potential,
the same which one expects in "cigar" model. Also the potential 
between two fast decaying D0-branes is calculated and the
general results is found in agreement with phenomenology.
Discussions are presented finally.


\section{On D-branes}
\setcounter{equation}{0}
D$p$-branes are $p$ dimensional objects which are defined as
(hyper)surfaces which can trap the ends of strings \cite{Po2}. 
One of the most interesting aspects of D-brane dynamics appear in their
{\it coincident limit}. In the case of coinciding $N$ D$p$-branes in a
(super)string theory, their dynamics is captured by a 
dimensionally reduced $U(N)$ (S)YM theory from (9)25+1 to 
$p+1$ dimensions of D$p$-brane world-volume \cite{W,Po2,Tay}. 
Due to matrix models \cite{BFSS,IKKT} it has been understood that 
the supersymmetric gauge theories corresponding to $p=0,-1$ contain many
aspects which one expects from 11 and 10 dimensional supergravities 
(see e.g. \cite{Ba}).

In case of D0-branes $p=0$, the above dynamics reduces to quantum mechanics
of matrices, because, only time exists in the world-line. The bosonic part
of the corresponding Lagrangian is \cite{BFSS, KPSDF} 
\bea\label{1.1}
L=m_0  \tr\; \biggl(\haf  D_tX_i^2 -\;\cs\; \CV(X)\;\biggl) 
, \;i=1,\cdot\cdot\cdot,9, 
\eea 
where $\frac{1}{2\pi\alpha'}$ is the tension of the fundamental string and 
$D_t=\partial_t-iA_0$ acts as covariant derivative in the 0+1 dimensional
gauge theory.  
For $N$ D0-branes $X$'s are in adjoint representation of $U(N)$ and have the
usual expansion $X_i=x_{i(a)}\;T_{(a)},\;\;a=1,\cdot\cdot\cdot,N^2$. 
The potential $\CV(X)$ is 
\bea\label{1.5} 
\CV(X)=-\;[X_i,X_j]^2. 
\eea 
In fact (\ref{1.1}) is the result of the truncation of the string theory 
calculations in the so-called {\it "gauge theory limit"} defined by
\bea\label{1.10}
l_s &\rightarrow& 0,\nonumber\\
\frac{v}{l_s^2}&=&{\rm fixed},\nonumber\\
\frac{b}{l_s^2}&=&{\rm fixed},
\eea
which $v$ and $b$ are the relative velocities and distances relevant to 
the problem and $l_s$ is the string length.

Firstly let us search for D0-branes in the above Lagrangian. 

For each direction $i$, there are $N^2$ variables and not $N$ which one
expects for $N$ particles. Although there is 
an ansatz for the equations of motion
which restricts the $T_{(a)}$ basis to its $N$ dimensional Cartan
subalgebra. This ansatz causes vanishing the potential (\ref{1.5}) and one
finds the equations of motion for $N$ free particles. In this case the $U(N)$
symmetry is broken to $U(1)^N$ and the interpretation of $N$ remaining
variables as the classical (relative) positions of $N$ particles is
meaningful. The centre of mass of D0-branes is represented by the trace
of the $X$ matrices.

In the case of unbroken gauge symmetry, the $N^2-N$ non-Cartan elements have
a stringy interpretation, governing the dynamics of low lying 
oscillations of strings stretched
between D0-branes. Although the gauge transformations mix the entries
of matrices and the interpretation of positions 
for D0-branes remains obscure \cite{Ba},
 but even in this case the centre of mass is meaningful. So the
ambiguity about positions only comes back 
to the relative positions of D0-branes.

To calculate the effective potential between D0-branes one should
find the effective action around a classical configuration. 
This work can be done by integrating over the quantum fluctuations 
in a path integral. For the diagonal classical configurations, 
which we know them as classical representation of D0-branes, 
the quantum fluctuations which must be integrated are 
 the off diagonal entries. According to the above picture, 
this work is equivalent with integrating over the oscillations 
of the strings stretched between D-branes. Because here we are faced 
with a gauge theory, and our 
interests is the calculation around a classical field
configurations, so it is convenient to use the background 
field method \cite{Abb} for calculation the effective action. 

For calculation the effective action we write 
(\ref{1.1}) in $D$ space-time dimensions in the form (in the units 
$2\pi\alpha'=1$ and after the Wick rotation $t\rightarrow it$
and $A_0 \rightarrow -iA_0$)
\bea\label{1.20}
L&=&m_0\tr\left(\frac{1}{4}[X_\mu,X_\nu]^2\right),\;\;\;
\mu ,\nu =0,1,...,D-1,\nonumber\\ 
X_0&=&i \partial _t +A_0,\;\;\;\;S=\int L dt,
\eea
which $\mu$ and $\nu$ are summed with Euclidean metric.
The one-loop effective action of (\ref{1.20}) has been calculated 
several times (e.g. see the Appendix of \cite{IKKT} with
ignoring the fermions) and the result can be expressed as
\bea\label{1.25}  
(\int dt)\;V(X_\mu^{cl}) =\frac{1}{2}Trlog\bigg(
P_\lambda^2\delta_{\mu\nu}-2iF_{\mu\nu}\bigg)-
Trlog\bigg(P_\lambda^2\bigg), 
\eea       
which the second term is due to the ghosts associated with gauge
symmetry and 
$$
P_\nu*=[X_\mu^{cl},*],\;\; 
F_{\mu\nu}\;*=[f_{\mu\nu},*],\;\; 
f_{\mu\nu}=i[X_\mu^{cl},X_\nu^{cl}],
$$ 
and
\bea\label{1.30}
P_\lambda^2=-\partial_t^2+\sum_{i=1}^{D-1} P_i^2,
\eea
with the backgrounds $A_0^{cl}=0$. 

\section{Static potential}
\setcounter{equation}{0}
Here we calculate the potential between two D0-branes at rest.
The classical solution which represents two D0-branes 
in distance $r$ can be introduced as
\bea\label{3.1}
X_1^{cl}=\haf \left( \matrix{ r & 0 \cr 0 & -r } \right),\;\;
X_0^{cl}=i\partial_t \left( \matrix{ 1 & 0 \cr 0 & 1 } \right),\nonumber\\
A_0^{cl}=X_i^{cl}=0,\;\;\;i=2,...,D-1                          .
\eea
So one finds
\bea\label{3.5}
P_1=\frac{r}{2}\otimes \Sigma_3,\;\;P_0=i\partial_t\otimes 1_4,\;\;
P_i=0,\;\;\;i=2,...,D-1,
\eea
where $\Sigma_3$ is the adjoint representation of the third Pauli matrix
, $\Sigma_3*=[\sigma_3,*]$. The eigenvalues of $\Sigma_3$ 
are 0, 0, +2, -2. 

The operator $P_\lambda^2$ will be found to be
\bea\label{3.10}
P_\lambda^2=-\partial_t^2\otimes 1_4 + \frac{r^2}{4}\otimes \Sigma_3^2,
\eea
and the one-loop effective action can be computed 
\bea\label{a}
V(r) &=& (\frac{D}{2}-1)Trlog\bigg(P_\lambda^2\bigg) \nonumber\\
&=&-  \;2(\frac{D}{2}-1) \int_0^\infty 
\frac{ds}{s}\int_{-\infty}^{\infty} dk_0\; \e^{-s(k_0^2+r^2)}\nonumber\\
&~&+\; {\rm traces\; independent\; of\; } r,
\eea
where 2 is for the degeneracy in eigenvalue 4 of $\Sigma_3^2$, and
$k_0$ is for the eigenvalues of the operator $i\partial_t$.
In writing the second line we have used  
$$
\ln \bigg(\frac{u}{v}\bigg)=\int_0^\infty \frac{ds}{s} \; 
(\e^{-sv}-\e^{-su}).
$$
The integrations can be done and one finds
\bea\label{3.15}
V(r)&=&-\; 2(\frac{D}{2}-1) \int_0^\infty \frac{ds}{s}
(\frac{\pi}{s})^{\haf} \e^{-sr^2} \nonumber\\
&=&\;4 \pi (\frac{D}{2}-1)\; |r|\; + \; 
 \infty\; {\rm independent \;of\;} r.
\eea       
The linear potential is the same of "cigar"
model, see e.g. \cite{collins}. Also it is the same which is consistent 
with spin-mass Regge trajectories \cite{ansar}. By restoring the 
$\alpha'$ the potential will be found to be
\bea
V(r)=\;4 \pi (\frac{D}{2}-1)\; \frac{|r|}{2\pi\alpha'}
\eea       
which has the dimension $lenght^{-1}$. 
By comparison with Regge model 
one can have an estimation for $\alpha'$ \cite{ansar}.

It is assumed that the above potential causes pair production when 
quark and anti-quark are separated sufficiently far. The minimum 
distance sufficient for pair production depends on $\alpha'$ and 
the mass of the lightest quark. So although the linear behavior 
of the potential is not for infinite separations, but
absence of free quarks remains expectable. 
\vspace{.5cm}

{\bf Fast Decaying D0-branes:}

For two {\it fast decaying} D0-branes one can again calculate 
the above potential. This work can be done by putting for example
a Gauss function for $k_0$ in the (\ref{a}). 
This work is equivalent with restricting the eigenvalues of the operator
$i\partial_t$, with this in mind that eigenvalues  
of operators ($X,\;i\partial_t$, ...) represent the information
correspond to classical values of D0-branes
\footnote{The eigenvalues of $i\partial_t$ are different from their
quantum mechanical analogy which due to the Schrodinger's 
equation are energy.}.
So one finds
\bea
V(r)&=&-\;2(\frac{D}{2}-1) \int_0^\infty 
\frac{ds}{s}\int_{-\infty}^{\infty} dk_0\; 
\bigg(\frac{1}{\Delta} \e^{\frac{-(k_0-T)^2}{\Delta^2}}\bigg)
\e^{-s(k_0^2+r^2)}\nonumber\\
&\sim& r^{\xi},\;\;\;\;\;\;0<\xi<1,
\eea
a result consistent with the phenomenology of
heavy quarks \cite{ansar} which we know that their weak 
decay rate grows with $(mass)^5$. In the 
extreme limit $\Delta \rightarrow 0$
 which the two D0-branes see each other
"instantaneously" one can take them as two D(-1)-branes (: D-instantons).
The dynamics of D(-1)-branes are described by the action (\ref{1.20})
but instead of the taking $X_0$ as $i\partial_t$ one takes
$X_0$ as a matrix which its eigenvalues represent the
"time"s which D(-1)-branes occur.
A classical solution as
\bea
X_1^{cl}=\haf \left( \matrix{ r & 0 \cr 0 & -r } \right),\;
X_0^{cl}=\left( \matrix{ t_0 & 0 \cr 0 & t_0 } \right),
A_0^{cl}=X_i^{cl}=0,\;i=2,...,D-1     ,
\eea
represents two D(-1)-branes occurred in distance $r$ and time $t_0$.
The potential can be obtained easily
\bea
V(r)\sim -2(\frac{D}{2}-1) \int_0^\infty 
\frac{ds}{s}\; \e^{-s(r^2)}\;\sim\;\ln r,
\eea
which is consistent with phenomenology \cite{ansar}.


\section{Conclusion and Discussion}
In this letter we calculate the effective potential between two D0-branes
and the result appeared to be the known linear one. For two fast decaying 
quarks, which in the extreme limit see each other instantaneously we obtained
potentials in agreement with the phenomenology. As further push of our
calculations it is reasonable to calculate scattering amplitudes by these
potentials. 

{\it Why non-commutativity?}\\
Special relativity in a modern compact definition may be 
represented as follows:\\
{\sl A modification of space-time which to be prepared 
as a ground for the natural and theoretically consistent 
propagation of fields.}

So one learns that the space-time makes a $X_\mu$ 4-vector which behaves
like the gauge field $A_\mu$ 4-vector (spin 1) under the 
boost transformations. 

Also in this way Supersymmetry(SUSY) is a natural continuation of the special relativity 
program: Including spin $\frac{1}{2}$ sectors to the coordinates of space-time
 which are corresponded to the fermions of the nature.
This leads one to the space-time formulation of the SUSY theories. 

Now, what may be modified if nature has non-Abelian (non-commutative) 
gauge fields? In present nature non-Abelian gauge fields can not 
make spatially long coherent states; they are confined or too heavy. 
But the picture may be changed inside a hadron or very near of an 
electron. In fact recent developments of string theories sound this 
change and it is understood that non-commutative coordinates and
non-Abelian gauge fields are two sides of one coin, as is mentioned 
in Sec. 2.
The future theoretical research in this area may make clear the
relations.


\end{document}